\documentclass[12pt]{article}

\usepackage[utf8]{inputenc}
\usepackage{charter}
\usepackage{fullpage}
\usepackage{hyperref}
\usepackage{graphicx}
\usepackage{lipsum}
\usepackage[sort&compress,numbers]{natbib}
\usepackage{hyperref}
\hypersetup{hidelinks}
\usepackage[total={6.5in,9in},top=1in,headsep=0.1in,headheight=1in]{geometry}
\usepackage{wrapfig,lipsum,booktabs}

\newcommand\snowmass{
\begin{center}
  \rule[-0.2in]{\hsize}{0.01in}\\
  \rule{\hsize}{0.01in}\\
  \vskip 0.1in
  Submitted to the Proceedings of the US Community Study\\ 
  on the Future of Particle Physics (Snowmass 2021)\\
  \rule{\hsize}{0.01in}\\
  \rule[+0.8in]{\hsize}{0.01in}\\[-2em]
\end{center}
}

\usepackage[firstpage=true]{background}
\backgroundsetup{contents={\parbox{6.5in}{\snowmass}}, scale=1,placement=top,opacity=1,color=black,position={3.25in,1.2in}}

\usepackage{fancyhdr}
\fancypagestyle{plain}{%
  \fancyhf{}%
  \fancyhead[C]{}
  \fancyfoot[C]{\thepage}
}

\fancypagestyle{empty}{%
  \fancyhf{}%
  \fancyhead[C]{{\it Puzzling Excesses in Dark Matter Searches and How to Resolve Them}}
  \fancyfoot[C]{\thepage}
}
\title{Snowmass2021 Cosmic Frontier White Paper: Puzzling Excesses in Dark Matter Searches and How to Resolve Them
}
\date{}

\usepackage{authblk}

\author[1,2]{Rebecca K. Leane\thanks{Co-ordinator, rleane@slac.stanford.edu}}
\author[3]{Seodong Shin\thanks{Co-ordinator, sshin@jbnu.ac.kr}}
\author[4]{Liang Yang\thanks{Co-ordinator, liyang@physics.ucsd.edu}}
\author[4]{Govinda Adhikari}
\author[5]{Haider Alhazmi}
\author[6]{Tsuguo Aramaki}
\author[7]{Daniel Baxter}
\author[8]{Francesca Calore}
\author[9]{Regina Caputo}
\author[10]{Ilias Cholis}
\author[11,12]{Tansu Daylan}
\author[13]{Mattia Di Mauro}
\author[14]{Philip von Doetinchem}
\author[15]{Ke Han}
\author[16,17,18]{Dan Hooper}
\author[19,20]{Shunsaku Horiuchi}
\author[21]{Doojin Kim}
\author[22]{Kyoungchul Kong}
\author[23]{Rafael F. Lang}
\author[24,25]{Qing Lin}
\author[26]{Tim Linden}
\author[15,27,28]{Jianglai Liu}
\author[29]{Oscar Macias}
\author[30,31,32]{Siddharth Mishra-Sharma}
\author[33]{Alexander Murphy}
\author[3]{Meshkat Rajaee}
\author[34]{Nicholas L. Rodd}
\author[31]{Aditya Parikh}
\author[35]{Jong-Chul Park}
\author[36]{Maria Luisa Sarsa}
\author[18]{Evan Shockley}
\author[32]{Tracy R. Slatyer}
\author[20]{Volodymyr Takhistov}
\author[37]{Felix Wagner}
\author[38]{Jingqiang Ye}
\author[39]{Gabrijela Zaharijas}
\author[18]{Yi-Ming Zhong}
\author[15]{Ning Zhou}
\author[40]{Xiaopeng Zhou}

\affil[1]{SLAC National Accelerator Laboratory, Stanford University, Stanford, CA 94039, USA}
\affil[2]{Kavli Institute for Particle Astrophysics and Cosmology, Stanford University, Stanford, CA 94039, USA}
\affil[3]{Department of Physics, Jeonbuk National University, Jeonju, Jeonbuk 54896, Republic of Korea}
\affil[4]{Department of Physics, University of California San Diego, La Jolla, CA 92093, USA}
\affil[5]{Department of Physics, Jazan University, Jazan 45142, Saudi Arabia}
\affil[6]{Department of Physics, Northeastern University, Boston, MA 02115, USA}
\affil[7]{Fermi National Accelerator Laboratory, Batavia, IL 60510, USA}
\affil[8]{Laboratoire d'Annecy-le-Vieux de Physique Théorique (LAPTh), USMB, CNRS,  F-74940 Annecy, France}
\affil[9]{NASA Goddard Space Flight Center, Greenbelt, MD 20771, USA}
\affil[10]{Department of Physics, Oakland University, Rochester, MI 48309, USA}
\affil[11]{Department of Physics and Kavli Institute for Astrophysics and Space Research, MIT, Cambridge, MA 02139, USA}
\affil[12]{Department of Astrophysical Sciences, Princeton University, Peyton Hall, Princeton, NJ 08544}
\affil[13]{Istituto Nazionale di Fisica Nucleare, via P. Giuria, 1, 10125 Torino, Italy}
\affil[14]{Department of Physics and Astronomy, University of Hawaii at Manoa, Honolulu, HI 96822, USA}
\affil[15]{School of Physics and Astronomy, Shanghai Jiao Tong University, Shanghai 200240, China}
\affil[16]{Theoretical Astrophysics Department, Fermi National Accelerator Laboratory, Batavia, Illinois, 60510, USA}
\affil[17]{
Department of Astronomy and Astrophysics, University of Chicago, 5640 South Ellis Ave., Chicago, IL 60637}
\affil[18]{Kavli Institute for Cosmological Physics, The University of Chicago, Chicago, IL 60637, USA}
\affil[19]{Center for Neutrino Physics, Department of Physics, Virginia Tech, Blacksburg, VA 24061, USA}
\affil[20]{Kavli Institute for the Physics and Mathematics of the Universe\\The University of Tokyo, Kashiwa, Chiba 277-8583, Japan}
\affil[21]{Mitchell Institute for Fundamental Physics and Astronomy, Texas A\&M University, College Station, TX 77843, USA} 
\affil[22]{Department of Physics and Astronomy, University of Kansas, Lawrence, KS 66045, USA}
\affil[23]{Department of Physics and Astronomy, Purdue University, West Lafayette, IN 47906, USA} 
\affil[24]{State Key Laboratory of Particle Detection and Electronics, University of Science and Technology of China, Hefei 230026, China}
\affil[25]{Department of Modern Physics, University of Science and Technology of China, Hefei 230026, China}
\affil[26]{Stockholm University and The Oskar Klein Centre for Cosmoparticle Physics,  Alba Nova, 10691 Stockholm, Sweden}
\affil[27]{Tsung-Dao Lee Institute, Shanghai Jiao Tong University, Shanghai 200240, China}
\affil[28]{Shanghai Jiao Tong University Sichuan Research Institute, Chengdu 610213, China}
\affil[29]{GRAPPA, University of Amsterdam, Science Park 904, 1098 XH Amsterdam, The Netherlands}
\affil[30]{The NSF AI Institute for Artificial Intelligence and Fundamental Interactions}
\affil[31]{Department of Physics, Harvard University, Cambridge, MA 02138, USA}
\affil[32]{Center for Theoretical Physics, Massachusetts Institute of Technology, Cambridge, MA 02139, USA}
\affil[33]{SUPA, School of Physics \& Astronomy, University of Edinburgh, Edinburgh, EH9 3FD, UK}
\affil[34]{Theoretical Physics Department, CERN, 1 Esplanade des Particules, CH-1211 Geneva 23, Switzerland}
\affil[35]{Department of Physics, Chungnam National University, Daejeon 34134, Republic of Korea}
\affil[36]{Centro de Astropart\'{\i}culas y F\'{\i}sica de Altas Energ\'{\i}as (CAPA), Universidad de Zaragoza, Zaragoza 50009, Spain}
\affil[37]{Institute of High Energy Physics, Austrian Academy of Sciences, 1050 Vienna, Austria}
\affil[38]{Physics Department, Columbia University, New York, NY 10027, USA}
\affil[39]{Center for Astrophysics and Cosmology, University of Nova Gorica, Vipavska 13,
5000 Nova Gorica, Slovenia}
\affil[40]{School of Physics, Beihang University, Beijing 102206, China}

\begin{document}

\maketitle

\begin{abstract}

Intriguing signals with excesses over expected backgrounds have been observed in many astrophysical and terrestrial settings, which could potentially have a dark matter origin. Astrophysical excesses include the Galactic Center GeV gamma-ray excess detected by the Fermi Gamma-Ray Space Telescope, the AMS antiproton and positron excesses, and the 511 and 3.5 keV X-ray lines. Direct detection excesses include the DAMA/LIBRA annual modulation signal, the XENON1T excess, and low-threshold excesses in solid state detectors. We discuss avenues to resolve these excesses, with actions the field can take over the next several years.

\end{abstract}

\newpage
\tableofcontents
\newpage

\section{Introduction}

A number of indirect and direct dark matter experiments have observed excess signals above background over the years. They have provided tantalizing hints but no definitive proof of a dark matter discovery. Understanding the origin of these puzzling excesses is an important task for the community in the coming decade, as it will provide insights and guidance on the future direction of the field. In this solicited white paper, we summarize the status of the observed signal excesses from astrophysical observations and direct detection experiments, and discuss the efforts and prospects in resolving the puzzles.

\section{Astrophysical Signals}

\subsection{Galactic Center Gamma-Ray Excess (GCE)}
\textbf{\textit{Editor: Rebecca Leane}}\\
\textit{Contributors: Francesca Calore, Regina Caputo, Ilias Cholis, Tansu Daylan,  Mattia Di Mauro, Dan Hooper, Shunsaku Horiuchi, Rebecca Leane, Tim Linden, Oscar Macias, Siddharth Mishra-Sharma, Aditya Parikh,  Nicholas Rodd, Tracy Slatyer, Gabrijela Zaharijas, Yi-Ming Zhong}\\

An excess of GeV gamma rays from the Galactic Center has been definitively detected by the Fermi Large Area Telescope, ``Fermi-LAT". The leading explanations for this Galactic Center Excess (GCE) are a new population
of millisecond pulsars, or annihilating dark matter. Solving this problem is of pressing importance; we may
either find the first evidence of dark matter interactions with the Standard Model, or confirm the existence
of a new population of pulsars. We discuss the actions that can be taken to solve this problem over the next few years.

\subsubsection{Obtaining an Accurate Galactic Diffuse Emission Model}
We open with perhaps the most pressing issue for understanding the GCE -- the need to improve the Galactic diffuse emission model. This is the dominant source of photons in the GeV energy range observed by gamma-ray telescopes. It arises due to cosmic rays, accelerated from a variety of mechanisms, impacting regions of gas, dust, and starlight that are concentrated in the center of the Galaxy. Galactic diffuse emission must be understood before we can draw conclusions about the GCE. Currently, our best Galactic diffuse emission models cannot reproduce what is observed in the Fermi data to what is expected within Poisson noise. This is a substantial systematic we know exists in understanding the GCE, and we do not yet know what preferences for characteristics for the GCE will be obtained once we finally have a sufficiently good Galactic diffuse emission model.

Building up new Galactic diffuse emission models is a complicated task requiring new modeling techniques, fits to new multi-wavelength data, and substantial computing resources. One critical improvement will be increasing the resolution of gas maps. Most available gas maps in the literature assume circular orbits of interstellar gas, some amount of temporal stability, and certain tracers of only limited completeness and fidelity. The central molecular zone (CMZ) is particularly problematic to model. Separately, inverse Compton scattering (ICS) requires an improved understanding of: star-forming regions and the distribution and intensity of associated light; the propagation of leptons, which are susceptible to Galactic winds and other local phenomena; and the energetics, stability, and associated signals of transient injection \cite{Petrovic:2014uda,Cholis:2015dea,Morris:2020aa}. To this end, we must make use of (local) cosmic-ray observations from the Alpha Magnetic Spectrometer (AMS-02) on board the International Space Station, as well as broad multi-messenger observations from radio to MeV and TeV energies, which will constrain the ICS emission and disentangle its degeneracies with synchrotron.

Presently, these ingredients are converted to gamma-ray emission maps assuming cylindrical symmetry of cosmic ray diffusion, but resolving the mystery of the GCE calls for anisotropic, three-dimensional modeling of diffuse emission. Promising initial work~\cite{Porter_2017,Johannesson:2018bit} remains impeded by computational challenges. Hydrodynamic simulation of interstellar gas~\cite{Pohl:2008,Macias:2016nev,Macias:2019omb} provides a viable way forward to resolve the distribution of gas in a region where the gas orbits are highly non-circular.  Ultimately, these modeling and computational strides are urgently required to reduce the systematic Galactic diffuse emission uncertainties.

\subsubsection{Understanding the Spatial Morphology of the GCE}
\label{sec:morph}

Precision measurements of the spatial morphology of the GCE could disentangle the different alternative explanations that have been proposed. Advances in this line of research should go hand in hand with efforts to improve the precision of the Galactic diffuse emission models, as we cannot make any conclusions without accurately modeling other dominant components in the gamma-ray sky.  

Early analyses of the GCE~(e.g., Ref.~\cite{Daylan:2014rsa}) using Galactic diffuse emission models constructed with GALPROP found that spherically symmetric (steepened) NFW templates were preferred to ellipsoidal NFW templates oriented along some arbitrary direction. Recently, owing due the Galactic diffuse emission models being improved in complementary ways, varying morphology preferences have been reported.

Some recent studies have improved the Galactic diffuse emission GALPROP models, finding that a DM morphology is preferred to the Galactic bulge morphology. Ref.~\cite{Cholis:2021rpp} produced improved models by incorporating known uncertainties on cosmic-ray propagation parameters, using high-precision cosmic-ray observations from AMS-02 and Voyager I.
Ref.~\cite{DiMauro:2021raz} incorporated a new weighted likelihood, as well as using improved diffuse templates.

On the other hand, other studies have improved the diffuse emission modeling in other ways, instead finding that stellar bulge templates provide a better fit than the DM templates. These approaches include using improved models for the interstellar gas ~\cite{Macias:2016nev,Bartels:2017vsx,Macias:2019omb,Abazajian:2020tww,Calore:2021jvg}, as well as using spatially flexible fitting procedures~\cite{Storm:2017arh}. Ref.~\cite{Storm:2017arh} has produced the best fit to data so far, but still does not reproduce the data at the level of Poisson noise. The fit is also not based on physical models, and it is not clear if the large number of degrees of freedom corresponds to a physical mechanism capable of producing the Galactic diffuse emission contribution.

Further investigations with further improved Galactic diffuse emission models (as discussed in the subsection above) are required in order to robustly confirm the the morphology of the GCE. New fitting techniques, discussed in Sec.~\ref{sec:fitting}, can also help make more accurate conclusions. Once the morphology of the GCE is known with reliable accuracy, this will provide strong clues to the origin of the GCE. 

If the GCE is established to follow the bulge morphology, this would allow us to set strong constraints on DM, see e.g. Ref.~\cite{Abazajian:2020tww}. Furthermore, if the GCE is due to millisecond pulsars (MSPs), the spatial morphology of the signal could reveal the formation mechanisms of the MSPs. Assuming the primordial formation scenario for MSPs in the GC, studies~\cite{Ploeg:2021mrr,Gautam:2021wqn} have implemented state-of-the-art population synthesis codes to build synthetic populations of MSPs. The picture that is emerging from such efforts is that the MSPs responsible for the GCE should (approximately) trace the distribution of old stars in the Galactic bulge---a composite structure made up of a triaxial barlike structure extending a few kiloparsecs and a concentrated nuclear component in the inner $\sim 200$ pc of GC. Other simulation-based studies~\cite{Gnedin:2013cda,Brandt:2015ula} have posited that the GC MSPs could have been
the result of depositions from tidally disrupted globular clusters. In such models it is expected that the GC MSPs are spherically symmetric distributed. More realistic simulation studies (including stellar binary interactions, pulsar kicks, and mixed formation scenarios) will be needed in order to reduce the model uncertainties on the expected spatial morphology of the MSPs population.

If the GCE is due to DM, this would provide a direct handle of the DM distribution in our Galaxy; see the section below for discussion of modeling the DM density profile.

\subsubsection{Improving Models of Milky Way DM Density} The intensity of the DM annihilation signal is dependent on the DM density profile. Improving our understanding of the Milky Way DM density profile allows us to directly compare the preferred morphology of the GCE to our expectations for DM, and the suitability of the explanation.

The latest results from hydrodynamic simulations of galaxy formation seem to point towards a flattening of the DM profile in the inner Galaxy, with a less steep cusp than a standard NFW profile~\cite{Schaller:2015mua,Calore:2015oya}. However, the uncertainties are large at the GC, where $N$-body simulations are limited in resolution; further improvements in these simulations are needed. There are also more recent results making use of Gaia suggested the density may actually be even steeper than NFW cusp, see e.g. Ref.~\cite{2020gaia}. However, these are still very uncertain for the inner kpc or so.

 In general, it will be important to make detailed comparisons between the morphology of the GCE signal and the DM profiles predicted by cutting-edge hydrodynamical simulations and the latest observations.

\subsubsection{Improving Extended and Point-Like Templates}
Astrophysical emission components other than the semi-steady-state Galactic diffuse emission will be important for understanding the GCE. For example, the Fermi Bubbles are extended gamma-ray lobes that dominate emission at high latitudes and high energies. Their low-latitude extension, where they are potentially degenerate with the GCE, may be spectrally and morphologically distinct from their well-observed high-latitude component. Understanding the origin of the Fermi Bubbles will be critical to utilizing extended templates for these gamma rays in sky regions that are relevant for the GCE. Two other extended emission components that lie along the Galactic plane~\cite{Balaji:2018rwz}, and which have as yet undetected counterparts in other wavelengths, call for improved modeling as well.

Similarly, the Galactic stellar bulge must be modeled in greater fidelity before we can make final conclusions about the nature of the GCE. State-of-the-art bulge models were obtained using~\cite{Coleman:2019kax} VISTA Variables Via Lactea (VVV) data to study the population of Red Clump (RC) giants in the Galactic bulge. The SkyFACT algorithm~\cite{Storm:2017arh,Coleman:2019kax} has been used to obtain a non-parametric model of the spatial distribution of the RC giant stars in the Galactic bulge. These new (peanut-like) templates may provide a significantly better fit~\cite{Coleman:2019kax} to the data than the boxy bulge templates~\cite{Freudenreich:1998}.
As discussed above in Sec.~\ref{sec:morph}, some recent studies find a preference for a stellar bulge over spherically symmetric emission at the Galactic center, though some arrive at the opposite conclusion. Understanding the sensitivity of these results to fitting choices, systematically accounting for possible degeneracies with other emission including point sources and the Fermi Bubbles, and, ultimately, interpreting the implications for dark matter annihilation are of utmost importance, and it is important to test the Galactic bulge templates with current fitting techniques (discussed in more detail below). Dynamical evolutionary modeling of the bulge combined with population synthesis modeling of gamma-ray populations \cite{Gonthier:2018ymi,Ploeg:2020jeh} will constrain what astrophysical source classes can explain and help provide theoretical guidance on interpretation of gamma-ray detections.

Before assigning a final interpretation to the GCE, we must also understand in a data-driven way if the GCE itself is significantly asymmetric with respect to any spatial axes, as appears compatible with recent theoretical investigations~\cite{Leane:2020nmi,Leane:2020pfc}. 
Higher fidelity numerical simulations, using insights and constraints from the Gaia satellite, can be used to understand the allowed morphologies of a dark matter signal, for instance. Alternately, ideas from image processing can dissect the data in novel ways, allowing access to new aspects of the GCE without forward modeling.

Finally, other gamma-ray emission components such as isotropic emission and complete point source catalogs are also critical for understanding the GCE. These will principally improved from the observational perspective, but theoretical advances will need to consistently incorporate these data in their entirety.

\subsubsection{Understanding the Shape of the GCE Energy Spectrum}
The shape of the GCE gamma-ray energy spectrum can help reveal its origin. While the best-fit parameters for annihilating dark matter (DM) and millisecond pulsars (MSPs) are predicted to produce a compatible spectrum at $\sim$1 GeV or above, they largely disagree at lower energies. At these energies, Fermi-LAT's point spread function substantially degrades, introducing large systematics that obscure the low-energy part of the spectrum. To detect meaningful deviations at the low-energy end, new MeV gamma-ray telescopes are required. New telescopes such as eASTROGAM~\cite{DeAngelis:2017gra} and AMEGO~\cite{McEnery:2019tcm}, will be more sensitive to this part of the spectrum, and therefore may be able to differentiate the two hypotheses. The expected sensitivity to point sources with eASTROGAM is a factor of 2 better for extragalactic objects than 10 years of Fermi data, for a 1-year observation ($1.2 \times 10^{-12}$ erg/cm$^2$/s vs $2.8\times 10^{-12}$ erg/cm$^2$/s at 100 MeV)~\cite{DeAngelis:2017gra}. An additional GCE hypothesis, other than DM or MSPs, is a cosmic-ray outburst event. Compared to MSPs and DM, outburst activity from the GC can endure energy losses that soften the energy spectrum at further distances from the GC~\cite{Petrovic:2014uda, Cholis:2015dea}, producing a marked difference in the spatial dependence of the signal. While currently the spectrum seems invariant in its position and shape~\cite{Calore_2015}, disfavoring an outflow event, an important task is to reduce large systematic uncertainties. Lastly, if the GCE is produced by the stellar bulge rather than DM, more detailed spectral analyses would be needed to determine any remaining potential DM contributions.

\subsubsection{Understanding the GCE Pulsar Luminosity Function} 

If the GCE arises from MSPs, their luminosity function provides a handle on the number of expected GCE MSPs. It is important to understand if there is conflict between potential GCE MSPs with the MSP luminosity function of known pulsars in globular clusters or the disk~\cite{Hooper_2013,Cholis_2015,Ploeg_2017,Gonthier_2018,Zhong:2019ycb,ploeg2020comparing,Dinsmore:2021nip}. Wavelet studies have set constraints on the potential GCE pulsar luminosity function~\cite{Zhong:2019ycb}, requiring a very large number of new pulsars to explain the GCE. A better understanding of the total number of pulsars/MSPs in the Milky Way may set a strong bound on the luminosity function. The luminosity function can also be used to determine the number of expected detections in X-ray. While being spun-up by a stellar companion to become a MSP, MSPs exist for a time as a low-mass X-ray
binary (LMXB). If one expects a similar MSP birth for the GCE and the Milky Way’s globular cluster population, the ratio of MSPs to LMXBs should be similar. The number of LMXBs already detected in the GCE region can be used to estimate the size of the population of GCE MSPs, and the number of LMXBs has found to be severely too low compared to the required number of MSPs to explain the GCE~\cite{Haggard_2017,Zhong:2019ycb}. However, multiple MSP formation channels exist, leading to potentially different MSP populations in the GC and globular clusters. More detailed studies (dynamical evolution, population synthesis, etc) will be needed to shed light on the MSP populations. This includes in M31, where the LMXB population and its spatial extent can be measured better than the Milky Way~\cite{Abazajian:2012pn}.

\subsubsection{Fitting and Characterization Methods}
\label{sec:fitting}

While the GCE is detected at a very high statistical significance, the systematic uncertainty is large, deriving from the significant underlying uncertainties on the Galactic diffuse emission, extended sources, and point sources. Characterizing the GCE in the presence of these large systematic uncertainties is a crucial step for the near-term future.
One well-developed method for characterizing the excess is the non-Poissonian template fit (NPTF) \cite{Malyshev:2011zi,Lee:2014mza,Lee:2015fea}, but recent demonstration of bias in NPTF results has called into question some of the conclusions \cite{Leane:2019xiy,Leane:2020nmi,Leane:2020pfc}. While efforts to reduce the susceptibility of NPTF results to diffuse mismodeling have commenced \cite{Buschmann:2020adf,Chang:2019ars}, substantial additional theoretical effort will be required before we can draw final conclusions based on the NPTF. For example, the NPTF does not yet incorporate energy information; spectral information could potentially play a determining factor in how we interpret the NPTF results.

Wavelet-based approaches to the data \cite{Bartels:2015aea,McDermott:2015ydv,Balaji:2018rwz,Zhong:2019ycb} offer a different perspective on the GCE. These approaches seek to increase the signal-to-noise for a given GCE hypothesis \cite{Bartels:2015aea,Zhong:2019ycb} and/or reduce systematic background uncertainties \cite{McDermott:2015ydv,Balaji:2018rwz} at the cost of reducing statistical significance. With the continuous wavelet method and the 4FGL point source catalog~\cite{Fermi-LAT:2019yla}, Ref.~\cite{Zhong:2019ycb} shows that a millisecond pulsar population with a luminosity function described by a power law with a constant index across many decades in luminosity, once considered a leading alternative to dark matter annihilation \cite{Hooper:2010mq, Abazajian:2010zy, Hooper:2011ti, Abazajian:2012pn, Gordon:2013vta, Abazajian:2014fta}, is not a viable candidate to explain the GCE. Alternative luminosity functions have been examined in~\cite{Dinsmore:2021nip}. 
 
Given the advances and challenges listed above, it is timely to reconsider our fitting methodologies. New statistical methods are being developed that do not present with the same biases exhibited by the NPTF~\cite{Collin:2021ufc}. Probabilistic cataloging \cite{Portillo_2017,Feder_2020} provides a way to infer the positions of sub-threshold point sources (in contrast to the NPTF, which marginalizes them out) at the cost of a large-scale computational challenge. Extending fits to simultaneously utilize rich multi-wavelength data can constrain the origins of the GCE, especially given expected observational strides. Discrete wavelet methods have the potential to identify the angular scale associated with the GCE.

Methods that leverage recent developments in machine learning offer a further path to weigh in on the GCE. These methods, using convolutional neural networks perhaps extended using Bayesian deep learning and simulation-based inference techniques, have recently shown promise in overcoming some of the issues and computational bottlenecks associated with the application of traditional statistical methods to the GCE~\cite{Caron:2017udl,List:2020mzd,List:2021aer,Mishra-Sharma:2021oxe}. These methods aim to implicitly learn the full likelihood associated with the forward model of the Fermi gamma-ray data in the Galactic Center. This is in contrast to the NPTF, for example, which for computationally tractability considers a simplified description of the data assuming each pixel to be statistically independent. By being able to account for pixel-to-pixel correlations, machine learning methods have been shown to respond more favorably to sources of signal and background misspecification~\cite{List:2020mzd,List:2021aer,Mishra-Sharma:2021oxe}.

By eschewing an approximate treatment of the PSF as in the case of the NPTF, machine learning methods can work with a finer spatial resolution~\cite{List:2021aer} and be extended to lower energies, offering additional discriminating power. The inclusion of energy binning information is also easily admitted, bypassing some of the computational challenges needed for the inclusion of spectral information in the case of the NPTF.

Other than their use for fitting, machine learning methods have also recently shown promise for point source identification~\cite{Panes:2021zig}, and could be used to better characterize the population of resolvable point sources with implications for the nature of the GCE.

\subsubsection{Detecting Pulsar Candidates in Other Wavelengths}
The GCE signal presents in GeV gamma rays. However, if the GCE is powered by MSPs, they may be detectable in other wavelengths. Some directions are:

\begin{itemize}
    \item \textit{Detecting pulsar candidates in radio--} If the GCE is powered by MSPs, they may also pulse into radio. This signal is challenging to find with traditional single-dish telescopes, such as the Greenbank Telescope. However, there are very good prospects with the already operating Very Large Array (VLA), and MeerKAT~\cite{Calore:2015bsx} as well as SKA in the future. If one uses the disk population to calibrate the bulge source modeling, then no or a few detections with already achievable sensitivity would imply that either there is no bulge population or that its radio properties (namely flux distribution) are substantially different from disk pulsars. There are however, complexities in interpreting a null observation of bulge pulsars. It is possible that all pulsars producing the GCE are radio quiet, although it is not clear how naturally can we expect that to happen. Improvement of radio-gamma pulsar theory, observation, and modeling is required. Furthermore, higher confidence in the point source methods and any localization would help interpret such searches.
    \item \textit{Detecting pulsar candidates with TeV-scale $\gamma$-ray telescopes--} Pulsar candidates may efficiently accelerate $e^{\pm}$ pairs. Evidence of this process is found in mild evidence for TeV halos around MSPs in HAWC observations (e.g., Ref.~\cite{Hooper:2018fih}), correlation between radio luminosity and far-infrared observation in star-forming-galaxies~\cite{Sudoh:2020hyu}, and high-energy tail of the GCE~\cite{Linden:2016rcf}. The $e^{\pm}$ pairs injected by a putative MSPs population in the GC could produce detectable TeV-scale inverse-Compton (IC) emissions. While prompt $\gamma$ rays from MSPs would trace the MSPs spatial distribution, the IC counterpart would exhibit an energy-dependent spatial morphology. The predicted IC spectra for MSPs distributed as the Galactic bulge vs NFW$^2$ profile are indistinguishable, but their spatial morphologies have recognizable features at TeV energies~\cite{Song:2019nrx}. Such differences may be used by future high-energy $\gamma$-ray detectors such as CTA to provide a viable TeV-scale handle for the MSP origin of the GCE~\cite{Macias:2021boz}. Due to gradual aging, MSPs may not be as bright as usually assumed. Increasing the number of required MSPs creates potential dynamical problems due to the increased mass budget of the required bulge MSP population~\cite{Hooper_2016}, though scenarios without disrupted globular clusters are also considered~\cite{ploeg2020comparing}.

\item \textit{Detecting pulsar candidates with X rays--} Recently, it has been shown that a large population of bulge MSPs is not excluded by current observations of compact X-ray sources detected by {\it Chandra} towards the Galactic center~\cite{Berteaud:2020zef}.
This very same approach also allows one to identify promising bulge MSP X-ray candidates. 
Dedicated follow-up campaigns and better X-ray spectral measurements are 
required to further reduce the number of MSP X-ray candidates and, eventually, detect the brightest objects.

\item \textit{Detecting pulsar candidates with gravitational waves--} Although quite distant in the future, 3rd generation GW ground-based telescopes have the potential to detect the cumulative signal from a population of bulge MSPs from the GC direction~\cite{Calore:2018sbp}, as dominating contribution to the Galactic stochastic GW background. For the time being,  analysis of already available data can set (not yet  competitive) limits on pulsar ellipticity.

\end{itemize}

\subsubsection{Finding a Consistent DM Signal Elsewhere}
To corroborate a potential DM explanation of the GCE, a signal consistent with DM needs to appear in other experiments and observables. Some targets are:

\begin{itemize}
    \item \textit{Dark matter in dwarf spheroidal galaxies--} 
 Dwarf Spheroidal Galaxies are very DM dense environments, with low $\gamma$-ray background, making them ideal targets for DM annihilation searches. Currently, no conclusive signal is seen in dwarfs, though the limits that arise here are consistent with a DM signal from the GCE~\cite{Albert_2017}. Systematics in background estimation at the dwarf position are traditionally not taken into account, but worsen the limits by another factor 2-3~\cite{Calore:2018sdx}. More recently, more substantial systematic issues have been pointed out which are important for understanding potential DM signals in dwarfs~\cite{Chang:2020rem, Ando:2020yyk}. These are crucial to understand and accurately account for. 
 Recent strides in modeling the density profiles of classical dwarf-spheroidal galaxies (dSphs) \cite{Hayashi:2020jze} are important to extend to ultra-faint objects (more of which are discovered all the time \cite{2019ARA&A..57..375S}), which potentially have similarly high or higher $J$-factors for dark matter annihilation \cite{Boddy:2019qak}.
 
 While Fermi will not obtain much improved results due to statistics, improvements are expected by finding more dwarfs with DES and Rubin, allowing a significantly increased sample. Radio and X-rays can also set limits on annihilating DM, especially from dwarfs, though these can depend on magnetic field structure of targets, and sizeable systematics need to be improved~\cite{Jeltema_2008,Chan_2017}. 
 
 \item \textit{Dark matter in Andromeda--} 
An extended excess of gamma rays has been detected toward Andromeda (M31) \cite{Karwin_2019}. This signal is potentially consistent with the GCE, however, its interpretation is complicated primarily by the difficult to model MW foreground. For Fermi-LAT, the limited effective area and poor angular resolution $<\sim$GeV are also an issue \cite{Eckner:2017oul}. CTA will observe M31 and might detect high energy ($>\sim$ 50 GeV) counterpart from the M31 bulge, but it is not guaranteed since it could be too faint/too extended. New generation instruments such as e-ASTROGAM or AMEGO might give further clues on the origin of this emission and its possible common features with the GCE.    

\item \textit{Antiproton and antinuclei signals--} 
An excess has been identified in the spectrum of  cosmic-ray antiprotons at energies of $5-20$ GeV \cite{Cuoco:2019kuu, Cholis:2019ejx}. Intriguingly the range of dark matter models accommodating the antiproton excess is similar to those which could generate the excess of GeV-scale gamma rays~\cite{Cuoco:2019kuu, Cholis:2019ejx,Hooper:2019xss}, even though these two indirect dark matter probes are sensitive to different systematic uncertainties. An additional excess has been identified, in AMS antinuclei events~\cite{Cholis:2020twh}. Ongoing AMS-02 and future GAPS ~\cite{Aramaki2015} antinuclei searches can inform us about the possible DM mass, annihilation channel and cross section in association to both the CR antiprotons GeV excess and the GCE~\cite{Cholis:2020twh}.
 
\item {\it Other Wavelengths --} Dark matter annihilation would produce a population of energetic $e^+e^-$ pairs, which can emit synchrotron radiation~\cite{Linden:2011au, 2012syn}.

\end{itemize}

If the GCE is interpreted as a dark matter annihilation signal, then the observed excess is well-modelled by processes which produce $b\bar{b}$ or $\mu^{+}\mu^{-}$ final states~\cite{DiMauro:2021qcf}. Compelling explanations, which are both minimal and viable, can be found in thermal relic dark matter models with a Higgs portal coupling~\cite{Carena:2019pwq,Fraser:2020dpy}. Some such models can be probed at direct detection or collider experiments, settling complementary constraints on the possibilities.

It is crucial to note that while particle models alone are not enough to resolve the origin of the GCE, they typically predict other discovery channels. Keeping a set of compelling benchmark models in mind will help us establish a collection of potential complementary signals. Being cognizant of this broader set of constraints and looking at the overall picture is extremely valuable in determining the origin of the GCE signal and testing possible dark matter hypotheses. 

\subsection{AMS Antiproton Excess}
\textbf{\textit{Editor: Ilias Cholis}}\\
\textit{Contributors: Tsuguo Aramaki, Francesca Calore, Ilias Cholis,  Mattia Di Mauro, Philip von Doetinchem, Dan Hooper, Rebecca Leane}

\subsubsection{Status of the excess}

An excess at $\sim  10$ GeV energy in the cosmic-ray antiprotons flux observed by AMS-02 has been first claimed by \cite{Cuoco:2016eej, Cui:2016ppb}. This excess is now called the antiproton excess. Interestingly, the antiproton excess if it is a signal of dark matter it suggests similar properties for the dark matter mass, annihilation cross section and annihilation channels to those required to explain the GCE \cite{Cuoco:2016eej, Cuoco:2017rxb, Cholis:2019ejx, Cuoco:2019kuu}. 
Other early analyses stated that the uncertainties related to the production and propagation of antiprotons in the Milky Way make it difficult to claim the presence of any such excess \cite{Bringmann:2014lpa,Giesen:2015ufa,Kappl:2015bqa,Winkler:2017xor}. However, more recent analyses relying on improvements on cosmic-ray propagation, have found the antiproton excess to have a {\it local} significance of about $3-5$ $\sigma$  and be robust to i) the cross sectional uncertainties responsible for the production of antiprotons in inelastic nucleon-nucleon collisions, ii) the uncertainties of local cosmic-ray injection and propagation through the interstellar medium and iii) the effects associated to the time-, charge- and energy-dependent effects of cosmic-ray solar modulation \cite{Cholis:2019ejx, Cuoco:2019kuu}. However, these papers neglect error covariance, as the systematic correlation matrices are not released by AMS-02. With the aim to improve this scenario, very recently, Ref.~\cite{Boudaud:2019efq}, implemented their own error covariance estimates and found that the antiproton excess goes away. Ref.~\cite{Boudaud:2019efq} defined full energy-dependent correlations of the uncertainties due to  a benchmark transport model defined by AMS-02 B/C data~\cite{Genolini:2019ewc}, the antiproton production cross sections, and measurement effects quoted by the AMS-02 collaboration. The authors of \cite{Heisig:2020nse} reached a similar conclusion. Also, using the updated 7-year AMS-02 antiproton \cite{AMS:2021nhj}, the authors of \cite{DiMauro:2021qcf, Kahlhoefer:2021sha} reduce the significance of the antiproton excess further. The lack of published error correlation matrices by AMS-02 remains a substantial hurdle to understanding the robustness of this excess.

\subsubsection{Future Developments}

To finally establish the robustness of the antiproton excess against systematic uncertainties of instrumental nature, more collaboration between the experimental and theoretical communities is needed, to achieve a better understanding of the underlying correlations of the AMS-02 systematic instrumental errors is needed. In the future, the upcoming GAPS experiment~\cite{Aramaki2015} will measure with precision the antiproton spectrum in a low-energy region currently inaccessible to any experiment. With one flight, GAPS is expected to identify about $10^{3}$ antiprotons in the energy range $E < 0.25$\,GeV/n. The GAPS antiproton measurement will also allow for sensitive studies of systematic effects, in particular propagation of antinuclei in the interstellar medium and the heliosphere. 

In contrast with dark matter searches with antiprotons, which rely on small excesses on top of considerable astrophysical backgrounds, the unique strength of searches for cosmic antideuterons is their ultra-low astrophysical background~\cite{Donato:1999gy,Baer:2005tw,Donato:2008yx,Duperray:2005si,Ibarra:2012cc,Ibarra2013a,Fornengo:2013osa,Dal:2014nda, Korsmeier:2017xzj,Tomassetti:2017qjk,Lin:2018avl,Li:2018dxj,Cholis:2020twh}.
The production of antiprotons and heavier antinuclei can be strongly related. For instance, any dark-matter-induced signal in antideuterons should also find its imprint in the antiproton spectrum.
Over the last more than 20 years, it was pointed out many times that the first-time detection of low-energy cosmic antideuterons would be an unambiguous signal of new physics.

\begin{wrapfigure}{R}{.55\textwidth}
\vspace{-0.15in}
\centering
\includegraphics[width=.55\textwidth]{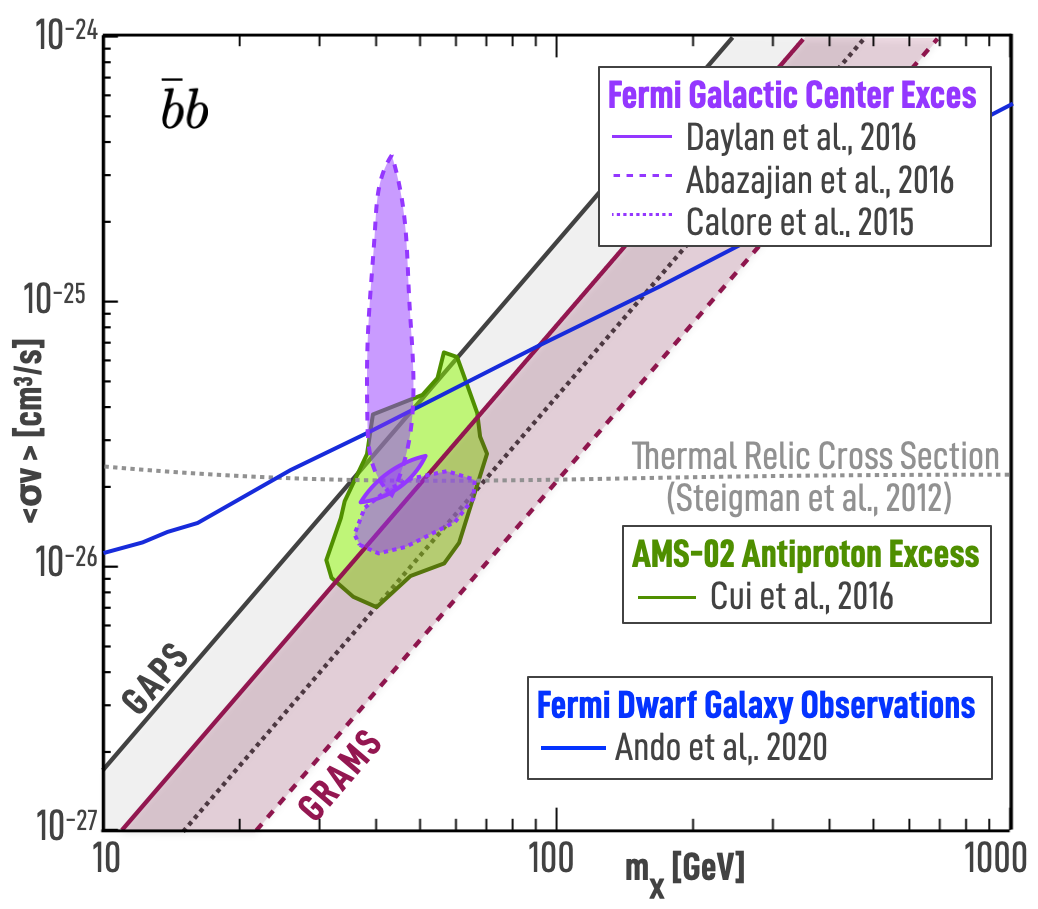}
\vspace{-0.3in}
\caption{{\it GAPS and GRAMS antideuteron sensitivities in the dark matte parameter space, along with the regions that could potentially explain the Fermi GCE and AMS-02 antiproton excesses \cite{aramaki2020dual}.}}
\label{fig:GAPS_GRAMS}
\vspace{-0.2in}
\end{wrapfigure}

GAPS that is optimized for low-energy antideuteron measurements, will be able to investigate the dark matter parameter space that could potentially explain the Fermi GCE and the AMS-02 antiproton excess. GRAMS (Gamma-Ray and AntiMatter Survey), a proposed mission beyond GAPS, with a further optimized detector with a LArTPC (liquid argon time projection chamber), will extensively explore the region in the parameter space, as seen in Figure \ref{fig:GAPS_GRAMS} \cite{aramaki2020dual}.

Furthermore, the AMS-02 collaboration has announced the remarkable observation of several candidate antihelium nuclei events~\cite{antihe,antihe2,antihe3}. 
This prompted significant public interest, and theoretical work.
Antihelium arriving from antimatter-dominated regions of the universe is already nearly excluded. Recently proposed models included modified antihelium formation models, dark matter annihilation, or emission from nearby antistars \cite{Blum:2017qnn,Tomassetti:2017qjk,Korsmeier:2017xzj, Poulin:2018wzu, Li:2018dxj, Cholis:2020twh}. 
Recent reviews can be found here ~\cite{Aramaki:2015pii,vonDoetinchem:2020vbj}.

Though these antihelium candidates are tentative, they require verification or refutation with either operational, upcoming, or completely new experiments. A positive signal would be genuinely transformative and refashion the field of cosmic-ray physics and potentially revolutionize our understanding of the Big Bang nucleosynthesis. AMS-02 will continue taking data for the remaining lifetime of the International Space Station, and GAPS will start its series of long-duration balloon flights soon. However, due to the ISS trajectory and experimental layout, AMS-02 focuses on a higher energy range than GAPS, which will fly with a low-geomagnetic cutoff trajectory from Antarctica. Furthermore, it is beneficial that both experiments use different identification techniques, reducing systematic uncertainties.

In addition, both the astrophysical propagation uncertainties and the antiproton production cross section uncertainties will need to be further reduced. The former can be reduced through cosmic-ray observations of multiple species. These observations will reduce the uncertainty on solar modulation's time-dependent effects and can be connected to observations of gamma-rays and lower-energy photons. 
The cross section for antiprotons to be produced by cosmic-ray interactions with the interstellar medium is key to interpreting cosmic antinuclei measurements.
The antiproton production cross section uncertainties in the energy range of AMS-02 are at the level of 10--20\%, with higher uncertainties for lower energies. For energies lower than the AMS-02 range, relevant for the GAPS experiment, a significant uncertainty on the source term from cross section normalization and shape exists. Future measurements at low center-of-mass energies ($<7$\,GeV), could improve these antiproton flux uncertainties~\cite{Donato:2017ywo}.
For heavier antinuclei made of multiple antinucleons, it is essential to note that every production process typically should produce antiprotons in much higher quantities. 
However, the heavier antinuclei formation processes are not well constrained~\cite{Gomez-Coral:2018yuk}.
In addition to the already available measurements \cite{Aamodt:2011zza,Aamodt:2011zj,Aduszkiewicz:2017sei, Aaij:2018svt}, more relevant antiproton production cross section measurements will be possible with ALICE, NA61/SHINE, and LHCb in the next years.
Important constraints for the antinuclei flux from dark matter annihilations are coming from the values of the diffusion coefficient, its rigidity dependence, and the Galactic halo size~\cite{Donato:2003xg}. Fits of cosmic-ray nuclei data for secondary-to-primary ratios are limited by uncertainties in the production  cross sections at the level of 10--20\%~\cite{0067-0049-144-1-153,2010A&A...516A..67M,2015A&A...580A...9G,Tomassetti:2017hbe,Genolini:2018ekk,Evoli:2019wwu}. Improvements in the accuracy of these production cross sections is important and will be possible, as with NA61/SHINE.

\subsection{AMS Positron Excess}
\textbf{\textit{Editor: Tim Linden}}\\
\textit{Contributors:  Mattia Di Mauro, Dan Hooper, Rebecca Leane, Tim Linden, Jong-Chul Park, Meshkat Rajaee, Seodong Shin}

\subsubsection{Status of the positron excesses at PAMELA and AMS-02}

For many decades, astrophysical positrons were thought to be primarily produced as ``secondaries" via the interactions of charged cosmic rays with interstellar gas. The detection of a hardening positron spectrum above 10~GeV -- first definitively detected by PAMELA~\cite{Adriani:2013uda}, and further verified by the Fermi-LAT (in an indirect way)~\cite{Ackermann_2012} and then with unprecedented precision by AMS-02 \cite{PhysRevLett.122.041102}, quickly rejected this hypothesis and posed a new challenge for astroparticle physics, which is denoted as the ``positron excess." 

Indeed, the flux ($\phi$) of very-high-energy positrons in units of $E^{3}\phi$ increases with energy, a result that is incompatible with pure secondary production (though see e.g.,~\cite{Blum:2013zsa}.) Several mechanisms have been explored in order to explain the new primary positron source, including the somewhat confusingly named ``secondary acceleration" of cosmic-ray positrons in supernova remnants. In this scenario the additional re-acceleration of positrons within the compact sources flatten the interstellar positron spectrum, making it appear like a primary source~\cite{2009PhRvL.103e1104B,Mertsch:2014poa}. 

A more exotic interpretation is associated to the annihilation or decay of dark matter particles in the Milky Way \cite{Bergstrom:2008gr, Chun:2008by, Ibarra:2009dr, Chun:2009zx, DiMauro:2015jxa, John:2021ugy}. However, the hypothesis that the positron excess is entirely explained by relatively conventional dark matter models is ruled out by the constraints obtained with other messengers such as $\gamma$ rays and antiprotons (see, e.g., \cite{DiMauro:2015jxa}). While early papers also focused on the possibility that ``leptophilic" dark matter models (where dark matter annihilates primarily to leptonic final states) may explain the properties of the excess while remaining consistent with $\gamma$-ray constraints~\cite{Arkani-Hamed:2008hhe}, recent analyses by Refs.~\cite{Bergstrom:2013jra,DiMauro:2015jxa,John:2021ugy} have utilized the smoothness of the rising positron spectrum to set severe constraints on the spiky positron spectra that would be produced by such leptophilic dark matter candidates. 

A more convincing explanation involves the production of primary e$^+$e$^-$ pairs in pulsar magnetospheres, along with their subsequent acceleration in the surrounding pulsar wind nebula. This interpretation has been significantly strengthened over the last few years. In particular, the Milagro Collaboration has reported the detection of $\gamma$-ray emission from 1-100 TeV from the direction of Geminga with an extension of $2.6^{\circ}$ \cite{2009ApJ...700L.127A}. Very recently, the HAWC and LHAASO Collaborations have reported the detection of extended $\gamma$-ray emission around three pulsars: Geminga, Monogem and PSR J0622+3749 \citep{Abeysekara:2017science,LHAASO:2021crt}. 
A very extended emission has been also detected in Fermi-LAT data above 10 GeV around Geminga demonstrating that these $\gamma$-ray halos could be structures detectable not only at TeV energies \cite{DiMauro:2019yvh}. These extended sources, called ``TeV halos" or ``Inverse Compton Scattering (ICS) halos" were first hypothesized (though were at the time deemed to be indetectable) almost 20 years ago by Ref.~\citep{Felix-book-2004vhec.book.....A} and later discussed in terms of the positron excess by~Ref.~\cite{Yuksel:2008rf, Hooper:2017gtd}. 

Notably, the same e$^+$e$^-$ pairs which upscatter interstellar radiation in order to produce the TeV halos also propagate to Earth where they can be observed as a hardened primary electron and positron spectrum. Moreover, the spectrum of the $\gamma$-ray emission from TeV halos can be used to calculate the expected e$^+$e$^-$ spectrum, with results that are in good agreement with models of the positron excess~\cite{Hooper:2017gtd}. In the last few years several publications followed this path.
For example Refs.~\cite{Hooper:2017gtd,Profumo:2018fmz,Tang:2018wyr,Fang:2018qco,DiMauro:2019yvh,DiMauro:2019hwn} found that the Geminga alone can produce an important part of the positron excess. 
Using this result Refs.~\cite{Hooper:2017gtd,Orusa:2021tts} demonstrated that the cumulative flux of positrons from Galactic pulsars can fit entirely the positron excess above 10 GeV with an efficiency for the conversion of spin-down luminosity into couples of electrons and positrons between a few up to ten $\%$.

A very intriguing consequence of the presence of these halos is that their extensions imply that the diffusion strength around pulsars is between 2-3 orders of magnitude smaller than the average value assumed for the propagation of cosmic rays in the Galaxy \citep{Hooper:2017gtd, Linden:2017vvb,  Abeysekara:2017science,Tang:2018wyr,Fang:2018qco,DiMauro:2019yvh,DiMauro:2019hwn,LHAASO:2021crt}. This evidence poses challenging theoretical questions on how an inhibited diffusion around astrophysical sources can be created.

In the last few years Refs.~\cite{Evoli:2018aza,Mukhopadhyay:2021dyh} published a model for which the inhibited diffusion is caused by the cosmic-ray gradient produced by the central source induces a streaming stability that self-confines the cosmic-ray population. In the most recent paper \cite{Mukhopadhyay:2021dyh}, where an error on the modeling of the ion neutral damping has been corrected with respect to \cite{Evoli:2018aza}, this model predict a suppressed diffusion for $\gamma$ ray energies below about 1 TeV. Instead, in the energies of interest for HAWC and LHAASO the diffusion coefficient is of the same order of the Galactic average.

An alternative explanation of the size of TeV halos has been investigated in \cite{Recchia:2021kty} where the authors have included the ballistic propagation that electrons and positrons encounter up to a timescale $\tau_c=3D(E)/c^2$ after injection. The CR transport is characterized by three regimes depending on the time $t$ after the injection: ballistic (for $t<< \tau_c$), diffusive (for $t > \tau_c$) and a transition between the two, that we call quasi-ballistic. The transition is governed by the energy-dependent mean free path $\lambda_c(E)$, which, for relativistic particles, is linked to the energy-dependent (as inferred both from theory and from the Galactic CR transport phenomenology isotropic diffusion coefficient through $D(E) =  \lambda_c(E)\, c/3$. A key potential discriminant between these two theoretical models is the measured efficiency of e$^+$e$^-$ acceleration, as the latter model requires approximately all of the pulsar spindown power to be converted into e$^+$e$^-$ pairs.

\subsubsection{Improvements Required}

Two areas of improvement are required to definitively determine the nature of the positron excess. The first includes modeling improvements necessary to determine the supernova remnant, pulsar, and dark matter contributions contributions to the positron flux. The second includes additional cosmic-ray and $\gamma$-ray observations which can significantly constrain the characteristics and origin of the excess. 

The precise modeling requirements depend on the scenario under consideration. For supernova remnant secondary acceleration, the main uncertainties are related to the production cross sections of electrons and positrons that is nowadays larger than at least 50\% \cite{Delahaye:2008ua}. In order to reduce these uncertainty estimations using the latest and precise cross section data from CERN experiments should be used. To improve the theoretical calculation for the positron flux from PWNe, we need to use future and precise measurements of HAWC, LHAASO and CTA to estimate the injection spectrum of positrons and the size of the low diffusion bubble present around sources. For dark matter models, we must examine models with non-conventional features, e.g., models with long-lived boosted particles in Refs.~\cite{Kim:2017qaw, Farzan:2019qdm}, and interpret these models within a robust framework that includes collider, indirect and direct detection constraints.

Critical observational improvements include advancements in the experimental sensitivity to the positron anisotropy. Models indicate that a sensitivity to a dipole anisotropy at the level of $10^{-4} - 10^{-3}$ would be required to discriminate between models powered by nearby sources (e.g., pulsar and supernova models) and potential dark matter explanations~\cite{Chu:2017vao}. Expected improvements in the sensitivity of existing water-Cherenkov telescopes (e.g., HAWC and LHAASO) and upcoming imaging atmospheric Cherenkov telescopes (e.g., CTA) will be capable of both constraining the population of TeV halos as well as placing improved $\gamma$-ray limits on dark matter indirect detection targets (e.g., dwarf spheroidal galaxies).

\subsubsection{Future prospects}

It is planned that AMS-02 will continue taking data until the end of the lifetime of the International Space Station. Continuing the measurements will allow extending the energy range of positrons towards higher energies and improve the accuracy of the spectrum. The upcoming Cherenkov Telescope Array (CTA) with an order of magnitude improvement in sensitivity over current telescopes will better detect, model and constrain TeV halo observations, and will also test a plethora of dark matter possibilities \cite{CTAConsortium:2017dvg}.

\subsection{511 keV Line}
\textbf{\textit{Editor: Seodong Shin}} \\
\textit{Contributors: Jong-Chul Park, Meshkat Rajaee, Shunsaku Horiuchi, Volodymyr Takhistov}

\subsubsection{Status of the 511 keV line}

The robust signal of 511 keV photon line originating from the decay of non-relativistic positronium has been observed from the galactic center for over 40 years~\cite{Siegert:2015knp, Knodlseder:2003sv, Jean:2003ci, Knodlseder:2005yq, Jean:2005af}.
The emission was first reported in 1972 by balloon-borne instruments and confirmed by several other experiments such as OSSE~\cite{Kinzer:2001ba} and more recently by the SPI spectrometer at INTEGRAL~\cite{Siegert:2015knp} (for a review see Refs.~\cite{Prantzos:2010wi, Teegarden:2004ct}).   
INTEGRAL/SPI has provided the most reliable imaging of the signal and revealed that most of the positrons are distributed in a nearly spherical region~\cite{Siegert:2015knp}.
Although astrophysical sources such as Pulsar~\cite{Wang:2005cqa}, X-ray binaries~\cite{Knodlseder:2005yq}, type Ia supernovae (SNIa)~\cite{Weidenspointner:2008zz, Kalemci:2006bz} and compact object mergers~\cite{Fuller:2018ttb} are not excluded, it is challenging to explain how positrons emitted by astrophysical sources could have a spherically symmetric morphology and a weak disk component. 
A spherical spatial distribution and the flux associated with the 511 keV emission could tantalizingly suggest a dark matter origin.

\subsubsection{Improvements required}

Further improvements in identifying the excess with dark matter scenarios can be made with various complementary studies in accelerators and other astrophysical observations.
Since conventional explanations with MeV-scale light dark matter particle annihilations~\cite{Boehm:2003bt, Boehm:2003ha, Huh:2007zw, Pospelov:2007mp} are now challenged by the constraints from delayed recombination~\cite{Wilkinson:2016gsy} and null results from dwarf galaxies~\cite{Siegert:2016ijv}, focuses should be given to other non-conventional explanations such as eXciting dark matter (XDM)~\cite{Finkbeiner:2007kk, Pospelov:2007xh, Cline:2010kv, Arkani-Hamed:2008hhe}, decaying DM~\cite{Picciotto:2004rp, Hooper:2004qf, Chun:2006ss}, pico-charged particles from dark matter decay~\cite{Farzan:2017hol, Farzan:2020llg} and PBH DM imploding neutron stars~\cite{Fuller:2017uyd}. 
The XDM possibilities can be tested in low-energy and high-intensity accelerators, which is beyond the scope of this white paper.
The last possibility with pico-charged particles can be tested by studying the correlation of the 511 keV emission with dwarf galaxies and halo magnetic fields.
Future observations will make it possible to study the magnetic field structure of dwarf galaxies more precisely~\cite{Beck:2013bxa}. Improved understanding of neutron star population distribution, abundance of heavy elements from r-process nucleosynthesis as well as kilonova observations will allow to further explore the scenario of PBH DM imploding neutron stars~\cite{Fuller:2017uyd}.

\subsubsection{Future Prospects}

Future prospects include more detailed understanding of spatial features (i.e. morphology) of the 511 keV signal.
Recently, reanalysis of the INTEGRAL/SPI data~\cite{Siegert:2021trw} has reported a departure from the   spatial morphology motivated by spherical DM distribution; instead, a preference for a correlation with the stellar distribution in the Milky Way bulge was found. Correlation with stellar distribution hints at the origin such as compact object mergers~\cite{Fuller:2018ttb}, instead of some alternatives such as DM annihilation or decay.
In addition, evidence was reported for a preference for a slight deviation from the stellar distribution, which may be due to kinematic kicks (e.g., supernova natal kicks) or propagation effects.
Further studies of these departures from sphericity, their robustness, and interpretations would help constrain whether the 511 keV signal is related to DM or not.
Moreover, proposed gamma-ray telescopes
such as AMIGO~\cite{Cirelli:2021fbo} or e-ASTROGAM~\cite{Tatischeff:2016ykb} which are aimed at  characterizing the diffuse MeV-scale gamma-ray emission from the halo of the Milky Way will help constraining the possible scenarios.

\subsection{3.5 keV Line}
\textbf{\textit{Editor: Nicholas Rodd}}\\
\textit{Contributors: Jong-Chul Park, Nicholas Rodd, Shunsaku Horiuchi, Volodymyr Takhistov}

\subsubsection{Status of the 3.5 keV line}

The 3.5 keV line is an anomalous X-ray line discovered in 2014 in both galaxy clusters and the Andromeda galaxy~\cite{Bulbul:2014sua,Boyarsky:2014jta}, using the {\it XMM-Newton} and {\it Chandra} telescopes.
While atomic transitions produce many lines at X-ray energies throughout the Universe, it is possible that the anomaly is not associated with any known process (although see Ref.~\cite{Jeltema:2014qfa}).
The hypothesis has been put forward that the line could be a signature of decaying DM, an exciting possibility is a predicted decay mode of sterile neutrino DM~\cite{Pal:1981rm,Abazajian:2001vt}. The sterile neutrinos, if produced through oscillations in the Early Universe~\cite{Shi:1998km}, would constitute warm dark matter (WDM)~\cite{Abazajian:2014gza} with the right level of matter suppression to explain small-scale structures of the Local Group (Milky Way and Andromeda galaxies) satellites~\cite{Horiuchi:2015qri}.
Intriguingly, depending on cosmological evolution and theoretical model, the putative 3.5 keV X-ray signal line could correspond to a sterile neutrino with a mixing large enough to be tested in upcoming laboratory experiments~\cite{Gelmini:2019esj,Gelmini:2019wfp}.
In the years following its discovery, there were results that both saw the line in additional objects~\cite{Urban:2014yda,Boyarsky:2014ska,Cappelluti:2017ywp}, and a number who searched for it but found no evidence of the emission~\cite{Horiuchi:2013noa,Malyshev:2014xqa,Anderson:2014tza,Tamura:2014mta,Jeltema:2015mee,Hitomi:2016mun,Gewering-Peine:2016yoj}, although given astrophysical uncertainties it was argued the results were not entirely inconsistent~\cite{Lovell:2018xng}.
(For a contemporaneous review, see Ref.~\cite{Abazajian:2017tcc}.)
Further, the possibility was put forward that a conventional astrophysical explanation in the form of charge exchange may explain the signal~\cite{Gu:2015gqm,Shah:2016efh}.

More recently, searches with significantly enhanced sensitivity have been devised that exploit the considerable archival data collected by {\it XMM-Newton}, exploiting the fact that every observation the instrument has made looks through a column density of the DM in the Milky Way~\cite{Dessert:2018qih,Dessert:2020hro,Foster:2021ngm}.
These analyses saw no evidence for an excess at 3.5 keV, and place the simplest DM interpretations of earlier detections under considerable tension, even under the most conservative assumptions for the DM distribution in the Milky Way.
A non-detection has also been reported from a large suite of archival {\it Chandra} data~\cite{Sicilian:2020glg}.
There has been discussion around the validity of these non-detections~\cite{Boyarsky:2020hqb} (although see the response in Ref.~\cite{Dessert:2020hro} and the public version of these analyses in Ref.~\cite{3p5-public-analysis}), suggesting that the field has not coalesced around a resolution for the anomaly.

\subsubsection{Future Prospects}

An observation with future instruments such as the XRISM telescope or Micro-X sounding rocket~\cite{XQC:2015mwy}, would likely lead the field to converge on the anomaly having or not having a simple dark-matter origin.

\section{Direct Detection}

\subsection{Annual modulation in Sodium Iodide}
\textbf{\textit{Editor: } Liang Yang}\\
\textit{Contributors: Govinda Adhikari, María Luisa Sarsa, Seodong Shin, Liang Yang}

\subsubsection{Annual Modulation signals of the DAMA experiment}
The DAMA/LIBRA (Large sodium Iodide Bulk for RAre processes) collaboration operates ~250 kg  ultra-low background NaI scintillators as dark matter detectors at the Gran Sasso National Laboratory (LNGS), Italy. It has consistently reported an excess of modulating low energy events between 2-6 keV$_{ee}$ region as the annual modulation signal of dark matter. Total accumulated data from DAMA/NaI (the prior generation experiment) and DAMA/LIBRA phase1+phase2 show a single-hit residual rate of (0.0103 $\pm$ 0.0008) cpd/kg/keV$_{ee}$, a measured phase of (145 $\pm$ 5) day, and a measured period of (0.9987 $\pm$ 0.0008) year with a significance of 12.9$\sigma$ \cite{DAMAPhase2}, consistent with expected modulation signal from dark matter particle interactions. The collaboration claims that no systematics or side reaction can mimic the annual modulation signal~\cite{DAMABkg}. The detectors were upgraded with higher quantum efficiency PMTs in 2012 for DAMA/LIBRA phase2 operation. The upgrade combined with new analysis techniques helped to lower the software threshold from 2 keV$_{ee}$ to 1 keV$_{ee}$. Phase2 data alone reports over 6 annual cycles corresponding to a total exposure of 1.13 ton$\cdot$yr and observes that the modulation signal persists with a significance of 9.5 $\sigma$ in the region of 1-6 keV$_{ee}$\cite{DAMAPhase2}. For the future, the collaboration plans to further reduce the analysis threshold and equip all PMTs with miniaturized low background preamplifier and improve the electronic chain with higher resolution digitizers~\cite{DAMA-2022}.

The DAMA claimed dark matter signal is controversial because other direct detection experiments with different target materials (Xe, Si, Ge) have failed to observe the signal based on standard WIMP models. Many theoretical attempts have been made to explain the discrepancy between different target materials but none has succeeded. Others have proposed modulating backgrounds as explanations of the observed signal including, among others, seasonal variation in muon flux \& modulation, scintillator phosphorescence effects, ambient temperature, and spallation neutrons from muons in the surrounding rock. DAMA has refuted and ruled out most proposed backgrounds but has not convinced the community that the observed modulation signal is due to dark matter. Independent experiments with the same target material NaI(Tl) as DAMA  would go a long way to resolve the controversy, allowing a model-independent test of the annual modulation signal. However, the lack of a deep understanding and/or modelling of quenching factors for scintillation of nuclear recoils in NaI(Tl) introduces some systematics in this test for DM candidates releasing the energy through nuclear recoils~\cite{NaI-quench-1,NaI-quench-2,NaI-quench-3,NaI-quench-4,NaI-quench-5,NaI-quench-6}. Better measurements of quenching factors for crystals having different properties using the same analysis methods and set-ups would help~\cite{NaI-quench-7}. Currently, two experiments, ANAIS at Canfranc Underground laboratory in Spain and COSINE-100 at Yangyang Underground laboratory in South Korea are running with Thallium-doped NaI detector as the target media and published their physics results \cite{ANAIS-AM, ANAIS-AM2, ANAIS-AM3, COSINE-AM, COSINE-AM2}. Other experiments including COSINE-200, SABRE, PICOLON are actively working on growing ultra-low background NaI(Tl) crystals to improve the experimental sensitivity \cite{COSINE200, SABARE, PICOLON}, while COSINUS uses NaI as scintillating bolometers to achieve event discrimination~\cite{COSINUS-2020}.

\subsubsection{Current status of ANAIS and COSINE-100}

\paragraph{ANAIS:}
ANAIS (annual modulation with NaI scintillators) is a dark matter direct detection experiment consisting of 112.5 kg of NaI(Tl) detectors in operation at the Canfranc Underground Laboratory (LSC), in Spain, since August 2017. ANAIS’ goal is to confirm or refute in a model independent way the DAMA/LIBRA positive result by studying the annual modulation in the low-energy detection rate. 

ANAIS-112 modules feature a very high light collection, at the level of 15 photoelectrons per keV in all nine modules. Another interesting feature is a Mylar window in the middle of one of the lateral faces of the detectors, which allows to calibrate with external sources of energies just few keV above the ROI for testing the DAMA/LIBRA result ([1–6] keV). Robust calibration down to the threshold is one of the assets of the experiment. On the other hand, considering altogether the nine ANAIS-112 modules, the average background in the ROI is 3.6 cpd/kg/keV after three years of data taking, while DAMA/LIBRA phase2 background is below 1 cpd/kg/keV. A full description of the experiment performance after the first year and a detailed background model can be found in Refs. \cite{ANAIS-performance, ANAIS-background} and an update after the third year in Ref. \cite{ANAIS-AM3}. 

ANAIS developed a blind analysis protocol in order to carry out unbiased annual modulation analysis in the region of interest, from 1 to 6 keV$_{ee}$. ANAIS-112 has presented results corresponding to 1.5, 2 and 3 years, following that protocol \cite{ANAIS-AM,ANAIS-AM2,ANAIS-AM3}. The three-year result, corresponding to an effective exposure of 313.95 kg$\cdot$yr, provides a best fit in the [1–6] keV ([2–6] keV) energy region for the modulation amplitude of -0.0034$\pm$0.0042 cpd/kg/keV (0.0003$\pm$0.0037 cpd/kg/keV). This result supports the absence of modulation in ANAIS-112 data, and it is incompatible with the DAMA/LIBRA result at 3.3 (2.6) $\sigma$, for a sensitivity of 2.5 (2.7) $\sigma$. The analysis takes into account the background contribution, without any subtraction, and including consistency checks and different modelling of the background time dependence \cite{ANAIS-AM3}. 

The statistical significance of ANAIS-112 result increases as expected and supports the prospects of reaching a sensitivity above 3 $\sigma$ in five years of operation~\cite{ANAIS-sensitivity}. The application of machine learning techniques to the ANAIS data analysis is expected to result in a relevant background reduction in the [1-2] keV energy region allowing for an increase in sensitivity with respect to the published prospects. The data corresponding to the three-year exposure are being reanalysed and results will be released in 2022.

\paragraph{COSINE-100:} 
The COSINE-100 experiment, located at Yangyang underground lab in South Korea, consists of 8 NaI(Tl) crystals with total mass of 106 kg. The crystals are submerged in $\sim$2000 L liquid scintillator that serves as an active background veto system. External backgrounds are further reduced by 3~cm copper and 20 cm of lead. Additionally, 37 muon counters are installed outside the lead shielding. Each crystal is optically coupled to two PMTs. Most of these crystals show high light output of $\sim$15 photoelectron/keV$_{ee}$. The experiment has been collecting physics data since September 2016.


Using 1.7 years of data, the collaboration has ruled out model-dependent dark matter interpretations of the DAMA signals in the specific context of standard halo model with the same NaI(Tl) target for various interaction hypotheses~\cite{COSINE-MD-result}. It also reported model independent searches of the annual modulation signal using 1.7 years data with 2 keV analysis threshold. The best fit for the 2–6 keV$_{ee}$ range has a modulation amplitude of 0.0092$\pm$0.0067 counts/keV/kg/day with a phase of 127.2$\pm$45.9 days. This data is consistent with both a null hypothesis and DAMA/LIBRA's 2-6 keV$_{ee}$ best fit value with 68$\%$ confidence level~\cite{COSINE-AM}. Recently, the collaboration lowered the analysis threshold from 2 keV to 1 keV, improved event selection, and gained more precise understanding of the detector background. With 2.82 yr livetime and 61.3 kg active mass, the collaboration reports best-fit values for the modulation amplitude of 0.0067 $\pm$ 0.0042 (0.0050 $\pm$ 0.0047) counts/(day$\cdot$kg$\cdot$keV) in the 1-6 (2-6) keV energy intervals with the phase fixed at 152.5 days. Again, the result is unable to distinguish between the DAMA observed modulation and no modulation~\cite{COSINE-AM2}. The detector will continue to operate until the end of 2022 when it will be replaced by the next phase of the experiment, COSINE-200.

\paragraph{Combined Analysis:} COSINE-100 and ANAIS collaborations have made good progress on probing the DAMA signal with the same target material. ANAIS-112 has refuted the DAMA positive modulation with almost 3$\sigma$ sensitivity in a model independent way with three years of data, and is expected to surpass 4$\sigma$ after completing six years of data taking (along with 2023 data), while COSINE-100 has excluded WIMPs as responsible of the DAMA/LIBRA signal in many scenarios. The two experiments have initiated discussions to conduct a combined analysis to search for annual modulation signals. The joint effort will not only have the best sensitivity to the DAMA signal in the near future, but will also allow better understanding of the detector backgrounds and collaboration on improving analysis techniques, thus providing guidance for future NaI experiments. 

\subsubsection{Future prospects}

\paragraph{ANAIS:}
ANAIS-112 will take data at least until August 2023, corresponding to six-year exposure. The application of the new machine learning analysis based on Boosted Decision Trees under development is expected to allow a sensitivity of 4$\sigma$ to DAMA/LIBRA result. To test DAMA/LIBRA beyond this level, other strategies are required. In the context of the ANAIS project, but beyond the timeline of ANAIS-112 data taking, there is ongoing R\&D aiming at operating the sodium iodide crystals at low temperature and replacing the PMTs by SiPMs. This approach offers several advantages: improvement of background budget and light collection, and reduction of the contribution of anomalous events attributed to the PMTs which at present reduce the experiment’s efficiency by requiring aggressive data selection protocols. This new detector concept combined with new more radiopure crystals could bring a high increase in sensitivity, both for low-mass WIMPs and testing of DAMA/LIBRA result.

\paragraph{COSINE-200:}
COSINE collaboration has been developing its own protocol for growing ultra-low background NaI crystals through powder purification, refining crystal growing and encapsulation techniques. It has successfully grown clean small size crystals (0.61-0.78 kg). Mass production of full size crystals is in progress with an expected background level of less than 0.5 counts/kg/day/keV which is lower than those of the DAMA crystals. COSINE-200, the next phase of the COSINE-100 experiment, is planned to start taking data in 2023 with these low background crystals at the Yemilab in South Korea. Besides providing more stringent tests of the DAMA results, COSINE-200 can perform low mass dark matter searches with the low background and higher light yield detectors. 

\paragraph{SABRE:} SABRE (Sodium Iodide with Active Background Rejection Experiment) collaboration is actively developing ultra-pure NaI crystal with the goal of achieving background in the energy region of interest of the order of 0.1 count/day/kg/keV, that is several times lower than the DAMA/LIBRA level. The crystals will be deployed in a liquid scintillator, which serves as an active anti-coincidence veto.  R\&D crystals have achieved an intrinsic background level comparable to DAMA ~\cite{SABARE}. The collaboration plans to deploy detectors in the Northern hemisphere at Laboratori Nazionali del Gran Sasso (LNGS), in Italy, and in the Southern hemisphere at the Stawell Underground Physics Laboratory (SUPL), in Australia. This simultaneous measurement will help disentangle any subtle effect due to cosmic muons, which have an opposite seasonal modulation in the two hemispheres.

\paragraph{PICOLON:} The PICOLON (Pure Inorganic Crystal Observatory for LOw-energy Neutr(al)ino) Collaboration has also focused its R\&D efforts on growing ultra-pure NaI crystals. It has successfully used a combination of recrystallization and ion exchange resins to reduce the $^{40}$K, $^{210}$Pb, and $^{226}$Ra backgrounds \cite{PICOLON}. The recent result of $^{210}$Pb in the NaI(Tl) is less than 5.7 $\mu$Bq/kg. The collaboration plans to start the dark matter search at the Kamioka Underground Laboratory with at least four NaI(Tl) scintillator modules, whose total mass is 23.4 kg, followed by phases II and III with total masses of 100 and 250 kg of NaI(Tl) crystal.

\paragraph{COSINUS:} The COSINUS (Cryogenic Observatory for SIgnatures seen in Next-generation Underground Searches) collaboration employs cryogenic calorimeter techniques to measure both the phonon and light signals from a NaI crystal, which allows discrimination of $\beta/\alpha$ events from the nuclear recoil events. The prototype detectors have achieved, ultra-low $^{40}$K background, light energy threshold of $\sim$ 0.6keV$_{ee}$ and phonon energy threshold of 5-6 keV~\cite{COSINUS-2018}. Additional R\&D is required to reach the goal of 1 keV for phonon threshold. The experiment with up to 50 kg of crystals will be deployed at Gran Sasso Underground Lab, Italy~\cite{COSINUS-2020}.

\subsection{XENON1T Electronic Recoil Excess}
\textbf{\textit{Editor:} Jingqiang Ye}\\
\textit{Contributors: Haider Alhazmi, Doojin Kim, KC Kong, Rafael F. Lang, Alexander Murphy, Jong-Chul Park, Seodong Shin, Evan Shockley, Jingqiang Ye, Liang Yang, Ning Zhou}

\subsubsection{Overview of the XENON1T Excess}
A low-energy electronic recoil (ER) excess below 7\,keV and most prominent between 2--3\,keV was observed in the XENON1T dark matter experiment~\cite{XENON:2020rca}. With a significance of $\sim3.5 \sigma$, this excess could be a statistical fluctuation, a hint of a new background process, or of new physics. With the more sensitive next-generation xenon experiments currently taking data, more insights into the nature of the XENON1T excess should soon be available. Here, we assume that the excess in XENON1T originates from physical events of some kind, either from a previously unmodeled background, or from physics beyond the Standard Model (SM). We first present potential backgrounds and how to confirm and/or reduce their presence, then briefly discuss a selection of potential explanations involving new physics, and conclude with an overview of the current status of the next-generation xenon experiments.

\subsubsection{Potential Backgrounds}

\paragraph{Tritium.}
Among all possible backgrounds, tritium is the most eye-catching candidate~\cite{XENON:2020rca, Robinson:2020gfu}. Tritium is a pure $\beta$ emitter with a Q-value of 18.6\,keV and its continuous energy spectrum peaks exactly between 2--3\,keV~\cite{Lucas_Unterweger}. The concentration of tritium required to explain the excess is extremely small, at around 3 atoms per kilogram of xenon. In general, two possibilities are considered for tritium to be introduced to an underground detector: 1) cosmogenic activation of detection media and detector materials above ground during fabrication, transportation, etc~\cite{Zhang:2016rlz}, and 2) its natural abundance in H$_2$O and H$_2$~\cite{PLASTINO200768}, i.e. tritiated impurities that can emanate from detector materials during operation. Since the half life of tritium (12.3\,years)
is longer than typical data taking time of a detector (which is of order of a couple years), tritium does not decrease significantly by its decay; however, it can be removed by a Hydrogen Removal Unit (HRU) during xenon purification~\cite{Dobi:2010ai}. 
A rigorous check of tritium hypothesis requires a more sensitive detector that comes with a larger exposure and a lower background level. A dedicated evacuation of a detector during commissioning is expected to reduce aforementioned tritiated impurities. In addition, it could be useful to study the ER rate change with different purification speeds that might change the equilibrium rate of tritiated impurities.

\paragraph{$^{37}$Ar.} Another potential background that should be highlighted is $^{37}$Ar~\cite{Szydagis:2020isq}, which decays via electron capture (EC) with a half life of 35\,days and can yield a 2.8\,keV peak~\cite{Barsanov:2007fe}. $^{37}$Ar is also possible to be in an underground detector by cosmogenic activation above ground~\cite{LUX-ZEPLIN:2022sad} and detector air leaks during operation~\cite{LUX:2015abn, XENON:2020rca}. Consequently, the cosmogenic activation of $^{37}$Ar should be taken into account for experiment planning, while regular measurements of $^{37}$Ar activity in the lab air are necessitated. $^{37}$Ar can be effectively removed by a cryogenic distillation system, which has been demonstrated by XENON1T~\cite{XENON:2021fkt} and PandaX-4T~\cite{Cui:2020bwf}. $^{37}$Ar was concluded not likely to explain the excess by the XENON collaboration, as its presence had been suppressed to a negligible level due to the long underground time of xenon and the underground cryogenic distillation before the science run started, as well as ruled out by the conservative estimation of detector leak and the measured $^{37}$Ar activity in the lab air~\cite{XENON:2020rca}.

\paragraph{Modeling of Known Backgrounds.}
There are also claims of potential systematic mismodeling in the XENON1T analysis, albeit mainly at much higher energies than the excess region~\cite{Dessert:2020vxy}. On the other hand, an independent analysis using the Noble Element Simulation Technique (NEST) largely confirmed the conclusion from XENON1T; i.e. that the excess was unlikely to originate from systematic effects~\cite{Szydagis:2020isq}. In addition, the spectral shape of $^{214}$Pb $\beta$ decay, the dominant background in the XENON1T experiment as well as the next-generation xenon experiments, is not precisely measured at low energies where some calculations suggest a possible increasing rate~\cite{XENON:2020rca,Haselschwardt:2020iey,Hayen:2020mod}. Last but not least, the $2\nu\beta\beta$ from $^{136}$Xe could have different spectral shapes as well~\cite{Kotila:2012zza}, which might alter the excess rate to some extent. It would be useful to measure these energy spectra with dedicated calibrations and/or analyses with the next generation experiments.

\paragraph{Other Backgrounds.} According to Ref.~\cite{Bhattacherjee:2020qmv}, additional backgrounds could be present in underground detectors due to cosmogenic production and activation during neutron calibration. Some backgrounds deposit low-energy peaks in the excess region, e.g. $^{41}$Ca (3.3\,keV, EC) and $^{49}$V (4.5\,keV, EC), while some $\beta$ emitters have broader spectra, e.g. $^{106}$Ru (39.4\,keV Q-value) and $^{137}$Cs (1176\,keV Q-value). However, it should be pointed out that the confirmation of any of those backgrounds requires more investigations, such as rigorous studies of the production rate and more importantly the removal by the purification system, as well as a rate consistency check from data.

\subsubsection{New Physics}

The XENON1T excess has been met with significant interest from the community, with 100 papers hitting the arXiv within the first month of its publication. Here we highlight a few candidates that might be consistent with the XENON1T excess.

\paragraph{Non-DM Candidates: Solar Axions, Bosonic Matter, Solar Neutrinos}
The energy range of the XENON1T excess could be consistent with absorption of a beyond-SM light particle produced in the Sun, for instance a new scalar, pseudoscalar, or vector species~\cite{Alonso-Alvarez:2020cdv, Bloch:2020uzh}. Being produced in the Sun, a detection of such particles would not necessarily shed light on the nature of dark matter, but would still be exciting markers of new physics. 

One potential new-physics explanation of the excess is the absorption of solar axions or axion-like particles (ALPs), a primary hypothesis considered by the XENON Collaboration~\cite{XENON:2020rca}. If these pseudoscalars exist, they could be produced in the Sun via their couplings to electrons, photons, and nucleons, and then absorbed in detection media by the axio-electric effect~\cite{XENON:2020rca} and inverse Primakoff effect\footnote{The original XENON analysis did not include the Inverse Primakoff effect.}~\cite{Gao:2020wer, Dent:2020jhf}. The XENON1T excess could be consistent with a solar axion/ALP signal with axion-electron coupling $g_{ae} \sim 3 \times 10^{-12}$; however, this is in strong tension with astrophysical analyses of stellar cooling and X-rays~\cite{Bloch:2020uzh}. Considering in particular the so-called QCD axion models, which would also resolve the strong CP problem in particle physics~\cite{Peccei:1977hh}, the limits on the axion couplings can be converted to constraints on the axion mass. In this case the XENON1T excess would be most consistent with a DFSZ-type axion~\cite{DINE1981199} with mass $O$(100) meV~\cite{XENON:2020rca}.

If the XENON1T excess persists in the next-generation xenon experiments described below, it should be possible to distinguish the solar axion hypothesis from other possible hypotheses, such as the tritium background, by the spectral shape difference~\cite{XENON:2020rca}. For a true confirmation of a possible axion signal, however, observing the same signal with a different technology would be required, e.g. the planned International Axion Observatory (IAXO)~\cite{Armengaud_2019}. Considering other bosons as well, future improvements can be made by complementary searches of such light particles in low energy and high luminosity accelerators. Clarification of various astrophysical processes constraining the new particle absorption possibilities can also guide the theoretical expectations. 

Another stellar explanation of the excess that has been considered is solar neutrinos with non-standard interactions. Perhaps the most straightforward such interaction arises from an enhanced neutrino magnetic moment, with the XENON1T result suggesting $\mu_{\nu} \sim 2 \times 10^{-11}~\mu_B$~\cite{XENON:2020rca}. Similarly to the solar axion hypothesis, this value for the neutrino magnetic moment is in strong tension with astrophysical constraints~\cite{Corsico_2014}. Moreover, many other models involving solar neutrinos and nonstandard interactions have been considered, e.g. other electromagnetic interactions, neutrino self-interactions,  active-to-sterile transition dipole moments, and light mediators, some of which can avoid existing constraints~\cite{Khan:2020vaf, Bally:2020yid, Shoemaker:2020kji}. The parameter space of fitting the excess in these models can affect cosmological and astrophysical observations. The existence of new neutrino interactions can be tested in various neutrino experiments throughout the world, including but not limited to GEMMA, CHARM-II, and Borexino~\cite{Boehm:2020ltd}.

\paragraph{DM Candidates: Boosted Dark Matter, Exothermic Dark Matter, Bosonic Dark Matter}

In general, the excess cannot be explained in terms of conventional, i.e., non-relativistic and single component, light WIMP {\it recoiling} electron~\cite{Kannike:2020agf}.
Therefore future improvements should be made by testing alternative explanations in dark sector theories beyond WIMP.
The first category is boosted dark matter (BDM) where the incoming light dark matter is boosted with relativistic energy by dark-sector structures~\cite{DEramo:2010keq, Belanger:2011ww, Agashe:2014yua, Bhattacharya:2014yha, Kong:2014mia,  Kopp:2015bfa, Kim:2016zjx, Giudice:2017zke, Heurtier:2019rkz}, astrophysical processes~\cite{Kouvaris:2015nsa, An:2017ojc, Emken:2017hnp, Calabrese:2021src}, scattering with charged cosmic rays~\cite{Bringmann:2018cvk, Ema:2018bih, Cappiello:2019qsw, Dent:2019krz, Bell:2021xff}, scattering with cosmic-ray neutrinos~\cite{Jho:2021rmn, Das:2021lcr, Chao:2021orr} or inelastic collision of cosmic rays with the atmosphere~\cite{Alvey:2019zaa, Su:2020zny} so that it can have enough kinetic energy inducing the recoil energy above 1\,keV at XENON1T~\cite{Fornal:2020npv, Su:2020zny, Cao:2020bwd, Jho:2020sku, Alhazmi:2020fju, Jho:2021rmn}. 
Interestingly, the first proposal of highly energetic electronic recoil by BDM at direct detection experiments including XENON1T was made in Ref.~\cite{Giudice:2017zke}, followed by the actual search in COSINE-100 in 2018~\cite{COSINE-100:2018ged}.
The second category is exothermic inelastic dark matter (XDM)~\cite{Graham:2010ca} where the incoming dark matter down-scatters the electron target producing a lighter dark matter component and transfers enough energy corresponding to the mass difference between the two dark matter components~\cite{Harigaya:2020ckz, Lee:2020wmh}.
The last category is bosonic dark matter where the incoming dark matter can be absorbed by detection media in the same way as bosonic matter produced in the Sun, including hidden photon dark matter, ALPs dark matter, etc.
Bosonic dark matter can induce a mono-energetic peak centered around its rest mass and the most favored mass by the excess is 2.3\,keV/$c^2$~\cite{XENON:2020rca}.

In all explanations, dedicated analyses of the atomic physics effects should be considered since the excess is observed at relatively low electronic recoil energy that is far below the electron mass.
These effects are due to the fact that target electrons are bound to the atom and in a state of interactions with other electrons as well as the nuclei. 
Therefore, the target electrons can no longer be treated as free at-rest particles and a theoretical improvement containing realistic treatments are required.
Adding to this, complementary searches of light DM at low energy and high luminosity accelerators, e,g, Ref.~\cite{Dutta:2019nbn}, are required.

\subsubsection{Status and Prospects of Next-generation Xenon Experiments}

\paragraph{XENONnT} As the upgraded version of XENON1T, XENONnT features a sensitive mass of 5.9\,tonne liquid xenon that is increased by a factor of 3 and a total ER background expected to be reduced by a factor of 6~\cite{XENON:2020kmp}. The activity of $^{222}$Rn that is the parent of the dominant $^{214}$Pb background was determined to be 4.2\,$\mu$Bq/kg in XENONnT through emanation measurements~\cite{XENON:2021mrg}. $^{222}$Rn was then decreased by a factor of $\sim$2 by the online radon column in XENONnT and is expected to be further reduced by another factor of 2 by switching into the final mode~\cite{ye:2021tap}. The xenon purity is also greatly improved by using the innovative liquid xenon purification system. The electron lifetime achieved in XENONnT is about one order of magnitude larger than the maximum drift time, meaning that almost all the liberated electrons from an interaction are able to drift upwards to the liquid surface without losing to the electronegative impurities~\cite{ye:2021tap}. With just a couple months of data, XENONnT is able to give more insights into the excess, in particular the preference between the solar axion hypothesis and the tritium hypothesis should the excess remains. A result is expected from XENONnT soon as the experiment is currently taking science data.

\paragraph{LZ} LUX-ZEPLIN (LZ) is a next-generation dark matter experiment using a 7\,tonne active mass of liquid xenon~\cite{LZ:2019sgr}. The expected low energy electronic recoil response of LZ to a variety of physics scenarios was recently presented in Ref.~\cite{LZ:2021xov}. Seven physical processes were considered: 1) an enhanced neutrino magnetic moment and 2) an effective neutrino millicharge, both for pp-chain solar neutrinos, 3) solar axions, 4) ALPs dark matter, 5) hidden photon dark matter, 6) mirror dark matter, and 7) leptophilic dark matter. Most of them have been proposed as possible explanations for the XENON1T excess. In all cases, significant progress over current sensitivity limits is expected. Moreover, the first few months of data from LZ should provide a rigorous test of the excess. The study also investigated the sensitivity dependence on the $^{222}$Rn level realized in actual data, which remains somewhat uncertain, showing that the dependence is fairly minimal. It also highlighted the impact on discovery sensitivity arising from ‘unexpected but possible’ backgrounds, specifically $^{37}$Ar and $^{3}$H. In a real experimental dataset, evidence for new physics would not be claimed if the observed excess were similarly consistent with some unexpected but possible background contamination. While in real data an externally-derived constraint on $^{37}$Ar and $^{3}$H concentrations or rates might be available, in this work a more conservative approach was taken in which no such constraint was assumed. This results in zero discovery sensitivity for signals with spectral shape identical to that of either $^{37}$ Ar or $^{3}$H, and sensitivity that is reduced for any signal with sufficiently high overlap with either of these two backgrounds.
Similarly to XENONnT, a result is expected from LZ soon. 

\paragraph{PandaX-4T} PandaX-4T is a dark matter experiment located at China Jinping Underground Laboratory (CJPL)~\cite{PandaX:2018wtu}. With a sensitive target of 3.7\,tonne liquid xenon, PandaX-4T reported the first dark matter search result using commissioning data of 0.63\,tonne$\cdot$year exposure and has placed the most stringent limit for WIMP-nucleon spin-independent cross section so far~\cite{PandaX-4T:2021bab}. However, PandaX-4T was not able to investigate the origin of this excess due to tritium leftover during a calibration for PandaX-II~\cite{PandaX-II:2020udv}, which is the predecessor of PandaX-4T. Due to the same reason, a similar ER search performed by PandaX-II after the observation of the excess only concluded that the XENON1T excess is within its constraints~\cite{PandaX-II:2020udv}. PandaX-4T is currently undertaking a tritium removal campaign and then restarts physics searches~\cite{PandaX-4T:2021bab}. The key issue to investigate the XENON1T excess is to have a robust measurement on the spectra of radon background at low energies. Dedicated calibrations and measurements on temporal variations under different purification conditions are planned at PandaX-4T, thus it is expected that an independent investigation on the XENON1T excess will be delivered from PandaX-4T in the near future.

\subsection{Solid State detectors (electron and phonon signals)}
\textbf{\textit{Editor:} Daniel Baxter}\\
\textit{Contributors: Daniel Baxter, Liang Yang, Jong-Chul Park, Felix Wagner}

Nearly all solid-state detectors currently operating with thresholds lower than 1~keV have observed statistically significant excesses of events that rise monotonically with decreasing energy, as one would expect from a dark matter signal~\cite{Proceedings:2022hmu}.
These excess rates have received considerably less attention, for example than the XENON excess, because both backgrounds and detector response in the sub-keV energy regime are far less well-understood, making systematic uncertainties large and often difficult to quantify. 
In fact, many of these excess rates are likely coming from new background sources and detector physics that was simply not observable with higher threshold devices. 
Here, we categorize these excess rates based on similar behavior and detector readout, along with plausible origins for each other than dark matter. 

\subsubsection{Low energy excess from Solid State detectors}

{\bf{Dark Rates}} 
Notably, over the last decade a number of technologies have developed (for the first time) sensitivity to single electron-hole pair creation~\cite{Tiffenberg:2017aac,SENSEI:2019ibb,SENSEI:2020dpa,SuperCDMS:2018mne,SuperCDMS:2020ymb}. 
This achievement substantially increases sensitivity to dark matter, but also to previously unexplored or unmodeled background processes that also give rise to individual electron-hole pair generation. 
All detectors are expected to have a fundamental dark rate based on thermal excitation of charges over the band gap of the material, but no detector has yet reached this limitation. 
Dark rate contributions can be broken into three categories: events accumulated during readout, events which scale with surface area exposure, and events which scale with bulk exposure~\cite{SENSEI:2021hcn}; the last of these is the most dangerous, as it precisely mimics the behavior of dark matter. 
It is also possible for dark rates to produce multi-electron events in an ionization detector, for example through event pile-up, but also through dark rate processes that are spatially correlated, such as from material defects. 
The best dark rates achieved come from silicon CCDs, with current limits on single(multiple) electron rates at 5(0.05) Hz/kg in SENSEI~\cite{SENSEI:2020dpa} and 7 Hz/kg in DAMIC~\cite{DAMIC:2019dcn}. \\

\noindent{\bf{Cherenkov Emission}} 
Recent work has shown that secondary emission from radiogenic backgrounds may cause rates that are peaked, similar to a dark matter signal~\cite{Du:2020ldo}. 
These backgrounds can mimic dark matter scattering in the low-energy regime and do not spectrally match predictions from direct radiogenic sources, like Compton scattering~\cite{Ramanathan:2017dfn,Botti:2022lkm}. \\

\noindent{\bf{Low-yield Detector Effects}} 
All cryogenic detector technologies either instrumented with TES or NTD sensor readout observe event excesses containing thousands of events with energies rising below a few hundred eV down to detector threshold~\cite{SuperCDMS:2020aus,EDELWEISS:2019vjv,EDELWEISS:2020fxc,CRESST:2019jnq}. 
While these excess rates far exceed the typical $5\sigma$ requirement for a detection, they cannot yet be taken as evidence for a dark matter signal as they are not quite consistent across detectors. 
More importantly, some experiments (in particular CRESST~\cite{Stahlberg:2021gqi}) observe a clearly decaying time dependence of the excess rates, which is in strong disagreement with the slightly modulating time dependency one would expect from a dark matter signal. 
Of the cryogenic detectors, CRESST-III observes by far the lowest rate in this regime using CaWO$_4$ crystals. 
\\

\noindent{\bf{Other}}
One solid-state detector excess does not fall into the categories listed above, specifically the excess observed above the 50~eV analysis threshold of DAMIC at SNOLAB~\cite{DAMIC:2020cut,DAMIC:2021crr}. 
This excess rate differs from those above in a number of ways.  
First, DAMIC has demonstrated that the excess events likely cannot be attributed to the known systematic uncertainties present in their background model, as they appear to be spatially consistent with events occurring in the bulk of the detector material~\cite{DAMIC:2021crr}. 
Second, the silicon CCDs used by DAMIC have been calibrated to have highly linear energy response all the way down to the analysis threshold of the excess~\cite{DAMIC:2016lrs}.  
Finally, the overall rate of the excess is quite low compared to other ionization detectors, consisting of only 17 events above 50~eV in 11~kg-days~\cite{DAMIC:2020cut}. \\

\subsubsection{Near and future prospects}
Significant advances in low-threshold detectors will allow further exploration of these excess event rates, along with robust calibration and modeling of detector backgrounds in this new energy regime. 
These efforts, already underway, will reveal whether a true dark matter signal could be hiding amongst the noise. 
For example, the application of Skipper CCDs by SENSEI-100, DAMIC-M~\cite{Castello-Mor:2020jhd,Settimo:2020cbq}, and Oscura~\cite{aguilararevalo2022oscura,brn} will thoroughly probe the viability of the DAMIC at SNOLAB excess rate by leveraging the lower threshold (single electron-hole pair, $\sim 4$~eV) of such devices while continuing to push to lower single-electron dark rates in silicon detectors. 
Meanwhile, advancements in cryogenic detectors, for example through the application of KIDs~\cite{Wen:2021ypr,Ramanathan:2021wjf} or GJJs~\cite{Kim:2020bwm} should enable lower thresholds for phonon or electron-recoil detection. 
Another approach would be the usage of identical detectors with identical readout but different target materials, which might whether the excess is a material effect. 
There are further efforts on the way to use specialized veto-systems, to test the dependence on external stress and surface contamination. 
Furthermore, the discovery of the sensitivity of quantum detectors, and in particular qubits~\cite{Vepsalainen:2020trd,Wilen:2020lgg,2021NatCo..12.2733C}, to ionization radiation opens a new frontier of lower threshold detector development, which will in turn provide a much stronger lever with which to explore these excess rates.

\section{Conclusions and Outlook}

Many tantalizing excesses have been reported across dark matter direct detection and indirect detection experiments.
We have discussed future directions for both the theoretical and experimental fronts in order to understand the origin of these excesses.

An GeV-scale excess at the heart of our Galaxy has been detected in gamma rays by the Fermi Large Area Telescope (Fermi-LAT). The statistical significance of this ``Galactic Center Excess" is well-established, but the field has still not converged on identifying its origin. Leading explanations are either weak-scale annihilating dark matter, or a new population of gamma-ray emitting pulsars. The key barrier to understanding the nature of the excess is obtaining an accurate model for the Galactic diffuse gamma-ray foreground -- this component makes up the bulk of the gamma rays in the region, but is not well understood. We extensively (and almost exhaustively) discuss the possible avenues to resolve this excess, from complementary discovery avenues, improved modeling avenues, and improved fitting and characterization methods.

An excess of GeV cosmic-ray antiprotons in the AMS-02 observations has been identified. If a signal of dark matter annihilation, it requires similar mass,  cross section and annihilation channels to those required to explain the GCE. To establish the robustness of this excess, the underlying correlations of the AMS-02 systematic errors are needed. The observation of antideuteron or antihelium nuclei by AMS-02 and future GAPS and GRAMS will be an unambiguous signal of new physics. Antiproton and antinuclei dark matter searches will benefit from future production cross section measurements from inelastic hadronic collisions at low center-of-mass energies, and from further reduction of the astrophysical cosmic-ray propagation modeling uncertainties. 

Observations of the rising positron fraction have sparked considerable intrigue over the past fifteen years due to their potential dark matter explanations. However, high-energy gamma-ray observations over the last few years have produced significant evidence that astrophysical e+e- acceleration by a population of high-energy pulsars is the most likely explanation for the excess positron flux observed at Earth. In closing this mystery, these TeV halos have added new questions, as their morphology indicates that the simple picture of isotropic and homogeneous particle diffusion throughout the Milky Way is violated on moderate scales. Understanding this new phenomenon may play an important role in dark matter indirect detection searches over the next decade, using both positrons as well as cosmic-ray and gamma-ray probes.

The 511 keV line from the decay of non-relativistic positrionium coming from the galactic center has been observed for over 40 years. In particular, the satellite telescope INTEGRAL/SPI has provided a nearly spherical morphology of the signal. Various non-conventional dark matter scenarios have been proposed to explain the excess, which can be complementarily tested in accelerators and astrophysical observations. More detailed understanding of the morphology of the signal is crucial in identifying the origin of the signal.

The 3.5 keV line is an anomaly detected in the X-ray band that has been interpreted as a possible hint of decaying dark-matter. First observed in the datasets of XMM-Newton and Chandra in 2014, the anomaly exhibited several properties expected of dark matter, although possible astrophysical explanations such as charge exchange or a potassium emission line were also discussed. Challenges to the simplest dark-matter interpretations have arisen from non-observations, particularly from the halo of the Milky Way. Future instruments such as XRISM have the potential to largely resolve the debate.

The DAMA/LIBRA experiment has observed an annual modulation signal in NaI(Tl) detectors with a significance of~12.9$\sigma$. The dark matter interpretation of the signal is incompatible with direct dark matter search experiments with other target materials (Xe, Si, Ge). Recent experiments (ANAIS, COSINE-100) using the same target material NaI(Tl) have put strong constraints on the DAMA dark matter results. Future experiments with ultra-pure crystals will be able to definitively test the annual modulation signal. 

The XENON1T dark matter experiment observed a low-energy electronic recoil excess, which is below 7 keV and mostly prominent between 2 and 3 keV. This excess could originate from unmodeled backgrounds or physics beyond the Standard Model. Several potential backgrounds are discussed and the methods to mitigate, confirm and reject them are also proposed, including tritium and $^{37}$Ar. A selection of new physics models are presented, such as solar axions, solar neutrinos, boosted dark matter. More insights into this excess are expected to be available soon from several next-generation xenon experiments that are taking data now.

Solid-state particle detectors measuring sub-keV energy depositions observe statistically-significant excess rates that must be understood in order to maximize the sensitivity to dark matter. Significant effort is underway in order to resolve the possible origins of these excess rates, which likely include single electron dark rates, secondary emission from radiogenic sources, and crystal cracking due to material stresses (e.g. clamping forces). While these difficult-to-model processes are likely responsible for the majority of these excess rates, it remains plausible that some of the unmodeled excess rates are from dark matter scattering.

\bibliographystyle{unsrt}
\bibliography{main.bib}

\end{document}